\documentclass[twocolumn,preprintnumbers,amsmath,amssymb,aps, pra]{revtex4}
\usepackage{graphicx}
\usepackage{dcolumn}
\usepackage{bm}
\usepackage[all]{xy}
\usepackage{indentfirst}
\usepackage{multirow}
\usepackage{mathrsfs}
\usepackage{bbm}
\usepackage{euscript}
\usepackage{extarrows}
\usepackage{epstopdf}
\usepackage{amsmath, amsthm, amssymb}
\usepackage[urlcolor=blue,citecolor=blue,linkcolor=red,colorlinks=true]{hyperref}
\usepackage{verbatim}

\newtheorem{theorem}{Theorem}
\newtheorem{prop}{Proposition}
\newtheorem{example}{Example}

\begin{document}

\title{Coherence and entropy complementarity relations of generalized wave-particle duality}

\author{Kang-Kang Yang}
\email{2220501004@cnu.edu.cn}
\affiliation{School of Mathematical Sciences, Capital Normal University, Beijing 100048, China}
\author{Zhi-Xi Wang}
\email{wangzhx@cnu.edu.cn}
\affiliation{School of Mathematical Sciences, Capital Normal University, Beijing 100048, China}
\author{Shao-Ming Fei}
\email{feishm@cnu.edu.cn}
\affiliation{School of Mathematical Sciences, Capital Normal University, Beijing 100048, China}

\begin{abstract}
The concept of wave-particle duality holds significant importance in the field of quantum mechanics, as it elucidates the dual nature encompassing both wave-like and particle-like properties exhibited by microscopic particles. In this paper, we construct generalized measures for the predictability and visibility of $n$-path interference fringes to quantify the wave and particle properties in quantum high-dimensional systems. By employing the Morozova-Chentsov function, we ascertain that the wave-particle relationship can be delineated by the average coherence. This function exhibits a close correlation with the metric-adjusted skew information, thereby we establish complementary relations between visibility, predictability, and quantum $f$ entropy, which reveals deep connections between wave-particle duality and other physical quantities. Through our methodology, diverse functions can be selected to yield corresponding complementary relationships.
\end{abstract}

\maketitle
\section{introduction}
The concept of wave-particle duality (WPD), as one of the fundamental principles in Bohr's complementarity theory, plays a pivotal role in quantum mechanics by elucidating the intrinsic disparity between the quantum realm and the classical domain \cite{Bohr1928}. Microscopic objects (such as photons, electrons, and even large organic molecules) exhibit both wave-like and particle-like behaviors when passing through an interferometer. However, these two behaviors cannot be observed simultaneously, leading to the relations of wave-particle duality. The quantitative analysis of this phenomenon was initially presented by Wootters and Zurek \cite{WotZrk} and has since been extensively studied by subsequent researchers \cite{Englert1996,Jaeger,GreenYas}. Notably, Englert proposed the following elegant duality relation \cite{Englert1996},
\begin{eqnarray}\label{WPDR1}
\mathcal{P}^2+\mathcal{V}^2\leq 1,
\end{eqnarray}
where $\mathcal{P}$ represents path information (predictability), while $\mathcal{V}$ denotes the visibility of interference fringes in a two-path interferometer. This trade-off relation imposes limitations on the information that can be simultaneously contained within both particle and wave aspects.

The quantitative investigation of WPD in $n$-path ($n\geq2$) interferometers was first proposed by D\"{u}rr \cite{Durr} who introduced a generalized predictability measure $P$ and a generalized visibility measure $V$. The former is determined on diagonal entries of the density matrix while the latter depends on non-diagonal elements. Subsequently, Englert et~al. \cite{Englert2008,Durr,TsuiKim,Peng,LuX} refined a reasonable criteria for these two quantifiers, which requires that normalization, invariance under relabeling, and convexity be satisfied. This framework is adopted in this study to quantify wave-particles by utilizing some functions.

With the development of quantum resource theory, there are growing interests in exploring the relationship between WPD and various quantum information concepts, such as entanglement \cite{Jakob1,TsuiKim,Jakob2}, coherence \cite{Bagan,Roy,LuoSun,BuKF,Bera}, entropic uncertainty \cite{Coles1,Coles2}, and quantum state discrimination \cite{LuX}. The predictability-visibility-concurrence triality relation has been experimentally and theoretically proven in Ref.\cite{Peng}. The equivalence between WPD and entropic uncertainty has been explored in \cite{Coles1}. The relationship between coherence and path information has been presented in Ref.\cite{Bagan}. The complementary relations between WPD and entangled monotones have been proposed and proven in \cite{Basso}. The complementary relationship between WPD and mixedness in $n$-path interferometers has been revealed in \cite{TsuiKim}. Sun and Luo \cite{SunLuo1} were the first to utilize coherence for quantifying interference from the perspective of Wigner-Yanase skew information. Regarding the quantification of coherence, there have been numerous recent research studies with corresponding results. It is worth mentioning that Sun et~al. \cite{Sun} have quantified coherence relative to channels using metric-adjusted skew information; building upon this, Fan et al. \cite{Fan} proposed an expression for average coherence while introducing a new entropy called quantum $f$ entropy. However, the idea of quantitatively studying WPD using metric-adjusted skew information has not yet been implemented in multipath interferometers, making it both novel and natural to investigate the complementary relations between WPD and quantum $f$ entropy.

In this paper, we study the quantitative relations between WPD and some quantum information measures, such as entropy and skew information. We first establish a generalized measure for quantifying the predictability and visibility properties of particles and waves in multi-path interferometers, utilizing a specialized symmetric operator concave function based on the spectral of density matrices. The function is closely associated with the metric-adjusted skew information, serving as a crucial link to establish the quantitative relation between predictability and visibility, as well as average coherence. We find that the sum of the generalized predictability and generalized visibility is less than or equal to one. Furthermore, complementary relations among predictability, visibility, and quantum $f$ entropy are revealed, and the trade-off relations are illustrated through detail examples.

\section{Measure of particle and wave aspects}
\subsection{Preliminaries}
An acceptable measure of path knowledge is a continuous function $P(\rho^{d})= P(\rho_{11},\rho_{22},\ldots,\rho_{nn})$ of the diagonal entries of a density matrix $\rho$, where $\rho^{d}$ stands for the diagonal part of $\rho$. As a valid predictability, the function $P$ should satisfy the following criteria:
\begin{equation}\nonumber
\parbox{0.78\columnwidth}{%
\makebox[0pt][r]
{{(1a)}~}$P=1$ iff $\rho_{ii}=1$ for one
$i$, i.e., the path is certain.\\
\makebox[0pt][r]
{{(2a)}~}$P=0$ iff $\rho_{ii}=1/n$ for all
$i$, i.e., the path is completely uncertain.\\
\makebox[0pt][r]
{{(3a)}~}$P$ is invariant under
permutations of the $n$ path labels.\\
\makebox[0pt][r]
{{(4a)}~}$P$ is convex, namely, for any two density
matrices $\rho_1$ and $\rho_2$, one has for $\rho=(1-\lambda)\rho_1+\lambda\rho_2$
($0\leq\lambda\leq1$),
\begin{displaymath}
P(\rho^{d})\leq(1-\lambda) P(\rho_{1}^{d})
+\lambda P(\rho_{2}^{d}).
\end{displaymath}
}
\end{equation}

Correspondingly, the wave aspect is characterized by the off-diagonal elements of $\rho$.
As a well defined measure of the wave aspect, the visibility $V(\rho)$ should satisfy the following conditions:
\begin{equation}\nonumber
\parbox{0.78\columnwidth}{
\makebox[0pt][r]{{(1b)}~}$V=0$ iff $\rho=\rho^{d}$.\\
\makebox[0pt][r]{{(2b)}~}$V=1$ iff $\rho=\sum_{j,k}|j\rangle \langle k|{\rm e}^{\mathrm{i}(\theta_{j}-\theta_{k})}/n$, i.e., $\rho$ is a pure state with equal diagonal elements.\\
\makebox[0pt][r]{{(3b)}~}$V$ is invariant under
permutations of the $n$ path labels.\\
\makebox[0pt][r]{{(4b)}~}$V$ is convex.
}
\end{equation}

Concerning the general measures of the predictability and the visibility, we denote by $\mathcal{F}_{op}$ the set of all functions $f:\left( 0,+\infty \right) \to \left( 0,+\infty \right)$ such that
i) $0\leq f(A) \leq f(B)$ for any $n\times n$ complex positive matrices $0\leq A \leq B$ ($f$ is operator monotone),
ii) $f(x)=xf(x^{-1})$ for all $x>0$ (symmetric),
iii) $f(1)=1$ (normalized),
iv) $f(0)>0$ (regular). For any $f\in \mathcal{F}_{op}$, we consider the Morozova-Chentsov function $c_{f}(x,y)=\frac{1}{yf(xy^{-1})}$, $x,y> 0$. The metric-adjusted skew information of $\rho$ with respect to an operator $H$ is defined by \cite{Hansen}
\begin{eqnarray}\label{quantum fisher in}
I_f(\rho,H)=\frac{f(0)}{2}\mathrm{tr}\{i[\rho,H]c_{f}(L_\rho, R_\rho)i[\rho,H]\},
\end{eqnarray}
where $[\rho,H]=\rho H-H\rho$ is the commutator of operators $\rho$ and $H$. Here $L_X(A)=XA$ and $R_X(A)=AX$ are left and right multiplication operators by $X$.
For $f\in \mathcal{F}_{op}$, denote
\begin{eqnarray}
\hat f(x)=\frac{1}{2}\left((x+1)-(x-1)^2 \frac{f(0)}{f(x)}\right),~ x>0.
\end{eqnarray}
The metric-adjusted skew information of $\rho$ can be further expressed as
\begin{small}
\begin{eqnarray}
I_f(\rho,H)=\frac{1}{2}\mathrm{tr}[\rho (H^\dag H+HH^\dag)]-\mathrm{tr}[H^\dag m_{\hat f}(L_\rho, R_\rho)(H)],
\end{eqnarray}
\end{small}
where $m_{\hat {f}}(x,y)=y\hat {f}(\frac{x}{y})=\frac{1}{c_{\hat {f}}(x,y)}$
is the corresponding generalized mean function and operator concave \cite{Pattrawut Chansangiam}.

Based on the properties of the metric-adjusted skew information, an average coherence measure $C_f$ with respect to operator monotone function $f\in \mathcal{F}_{op}$ has been has proposed in Ref.\cite{Fan}, which can be also interpreted as a measure of coherence for the depolarizing channel. For any $n$-dimensional state $\rho$ and $f\in \mathcal{F}_{op}$, the average coherence $C_{f}(\rho)$ is given by
\begin{eqnarray}\label{average cohe}
C_{f}(\rho)=\frac{n-\mathrm{tr}[m_{\hat f}(L_{\rho}, R_{\rho})]}{n+1},
\end{eqnarray}
where $\mathrm{tr}[m_{\hat f}(L_{\rho}, R_{\rho})]=\sum_{i,j}m_{\hat f}(\lambda_{i},\lambda_{j})$, and $\rho=\sum_{i}\lambda_{i}|i\rangle \langle i|$ is the spectral decomposition of $\rho$.

\subsection{Main results}
\begin{prop}\label{predic}
For any $n$-dimensional state $\rho$ and $f\in \mathcal{F}_{op}$, the predictability $P(\rho)$, \begin{eqnarray}\label{predictability}
P_{f}(\rho)=\frac{n-\mathrm{tr}[m_{\hat f}(L_{\rho^{d}}, R_{\rho^{d}})]}{n-1},
\end{eqnarray}
complies with the set of requirements $1a),2a),3a),4a)$.
\end{prop}

The proof is given in Appendix~\textbf{A}. Actually, the function $P(\rho)$ genuinely depends on the diagonal elements of $\rho$ and satisfies $0\leq P(\rho)\leq1$. $P(\rho)$ attains its maximal value 1 when there exists only one diagonal element 1, which is attained when $\rho$ is a pure state. This is consistent with the fact that the path is completely certain. When the diagonal entries of $\rho$ (i.e., the eigenvalues of $\rho^{d}$) are all equal, $P(\rho)$ vanishes and the path information is completely uncertain.
For the visibility we have the following conclusion, see Appendix~\textbf{B}.

\begin{prop}\label{visis}
For any $n$-dimensional state $\rho$ and $f\in \mathcal{F}_{op}$, the visibility $V(\rho)$ defined by,
\begin{eqnarray}\label{visibility}
V_{f}(\rho)=\frac{\mathrm{tr}[m_{\hat f}(L_{\rho^{d}}, R_{\rho^{d}})-m_{\hat f}(L_\rho, R_\rho)]}{n-1},
\end{eqnarray}
complies with the set of requirements $1b),2b),3b),4b)$, assuming that $V_{f}(\rho)$ does not increase under incoherent operations.
\end{prop}

It's easy to verify that $0\leq V(\rho)\leq1$. $V(\rho)$ reaches its minimum 0 when $\rho=\rho^{d}$ (i.e., the off-diagonal entries of $\rho$ are all 0). Meanwhile, $V(\rho)$ attains its maximal value 1 when $\rho$ is a pure state with equal diagonal entries, for which $\rho$ has non-zero off-diagonal elements. Namely, the function $V(\rho)$ characterizes the influence of the off-diagonal elements on the wave aspect. Consider the von Neumann measure $\Pi=\left\{\Pi_i=|i\rangle\langle i|\right.$: $i=1, \cdots, n\}$. $\Pi(\rho)=\sum_i \Pi_i \rho \Pi_i =\sum_i |i\rangle\langle i|\rho |i\rangle\langle i|=\rho^{d}$ denotes the full dephasing of $\rho$ in the computational basis $\{|i\rangle\}^{n}_{i=1}$. In the following we denote $V=V_{f}(\rho)-V_{f}(\Pi(\rho))=V_{f}(\rho)$ the change of $V_{f}$ due to the measurement. Additionally, it is easily seen that measurements do not alter the particle properties, i.e., $P_{f}(\rho)=P_{f}(\Pi(\rho))$.

Indeed, there are many ways to construct the predictability and visibility measures that comply with the requirements and satisfy certain duality relations. Different measures and trade-off relations characterize different aspects of the wave and particle properties.
Currently, these measures are mainly defined by either the elements or the spectra of a density matrix. The former requires comprehensive knowledge of the density matrix. For example, the visibility and predictability can be defined as functions of the off-diagonal and diagonal entries of the density matrix, respectively. In this paper, we adopt the latter approach to develop measures of predictability and visibility in multi-path interference without necessitating complete information about quantum states. In fact, the measures defined in these two ways are generally not equivalent. For instance, the measures with
predictability $P(\rho\mid\Pi)=\sum_{i=1}^n\langle i|\rho| i\rangle^2$ and visibility $V(\rho \mid \Pi)=\sum_{i \neq j}|\langle i|\rho| j\rangle|^2$ cannot be expressed in the form of measures defined in this paper, since $V(\rho \mid \Pi)\neq \operatorname{Tr}(\phi(\rho))-\operatorname{Tr}(\phi(\rho^d))$, where $\phi(x)=x^2$ \cite{TsuiKim}.

\section{Complementarity between WPD and information measures} From (\ref{average cohe}), (\ref{predictability}) and (\ref{visibility}) we have the following analytical relation among the predictability, visibility and the average coherence.

\begin{theorem}
For any state $\rho$ in $n$-dimensional quantum system and $f\in \mathcal{F}_{op}$, we have
\begin{eqnarray}\label{equ}
P(\rho)+V(\rho)=\frac{n+1}{n-1}C_{f}(\rho).
\end{eqnarray}
\end{theorem}

\begin{figure}
\begin{center}
\includegraphics[scale=0.6]{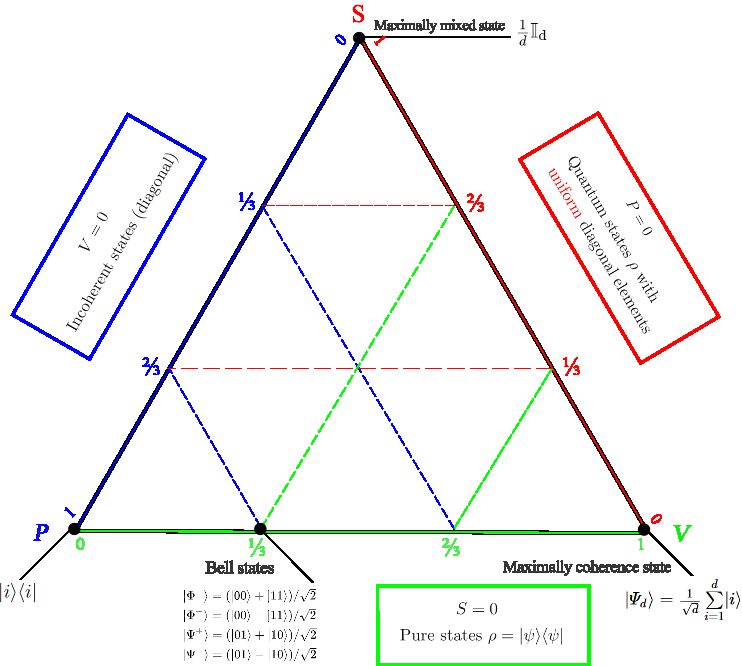}
\caption{The bottom solid green line corresponds to all pure states, $\textbf{S}=0$. The red solid edge on the right corresponds to the quantum states in which all diagonal elements are equal (the path information is completely uncertain), that is, $P=0$. The solid blue edge on the left corresponds to all diagonal states (incoherent states), i.e., $V=0$. The $\textbf{S}=\frac{1}{n-1}S_{f}$ with respect to $f=f_{WY}=((1+\sqrt{x}) / 2)^2$.
}\label{fig0}
\end{center}
\end{figure}
When $\rho$ is a pure state, one has $C_{f}(\rho)=\frac{n-1}{n+1}$ and (\ref{equ}) reduces to $P(\rho)+V(\rho)=1$. Associated to the Wigner-Yanase skew information, we take into account an operator monotone function $f(x)=f_{\mathrm{WY}}(x)=(\frac{\sqrt{x}+1}{2})^2$. Then we obtain correspondingly
\begin{eqnarray}
&&P_{f_{\mathrm{WY}}}(\rho)=\frac{n-(\operatorname{tr} \sqrt{\rho^d})^2}{n-1},\\
&&V_{f_{\mathrm{WY}}}(\rho)=\frac{(\operatorname{tr} \sqrt{\rho^d})^2-(\operatorname{tr} \sqrt{\rho})^2}{n-1},\\
&&C_{f_{\mathrm{WY}}}(\rho)=\frac{n-(\operatorname{tr} \sqrt{\rho})^2}{n+1},
\end{eqnarray}
which satisfy the trade-off relation (\ref{equ}). In particular, for two-dimensional quantum states, (\ref{equ}) reduces to the following relation,
\begin{eqnarray}\label{12}
P_{f_{\mathrm{WY}}}(\rho)+V_{f_{\mathrm{WY}}}(\rho)+S_{1 / 2}^2(\rho)=1,
\end{eqnarray}
where $S_{1 / 2}^2(\rho)=(\operatorname{tr} \sqrt{\rho})^2-1$ is the quantum unified-$(r,s)$ entropy, $S_r^s(\rho)=\frac{1}{(1-r) s}\left[\left(\operatorname{tr} \rho^r\right)^s-1\right]$ ($r \neq 1$, $s \neq 0$) for $r=1/2$ and $s=2$.

Note that (\ref{12}) can also be expressed in terms of the quantum Sharma-Mittal entropy $H_{q,r}(\rho)$ \cite{MDG},
$$
H_{q,r}(\rho)=\frac{1}{1-r}\left[\left(\sum_i\left(\lambda_i\right)^q
\right)^{\frac{1-r}{1-q}}-1\right],
$$
where $\lambda_i$s are the eigenvalues of $\rho$, $q$ and $r$ are two real numbers with $q>0$, $q \neq 1$ and $r \neq 1$. In terms of the quantum Sharma-Mittal entropy, (\ref{12}) has the following form,
\begin{eqnarray}
P_{f_{\mathrm{WY}}}(\rho)+V_{f_{\mathrm{WY}}}(\rho)+\frac{1}{n-1}H_{\frac{1}{2}, 0}(\rho)=1.
\end{eqnarray}

Based on the concept of average coherence, in \cite{Fan} a bona fide measure of entropy $S_{f}(\rho)$, named quantum $f$ entropy, has been recently proposed,
\begin{eqnarray}\label{f entropy}
S_{f}(\rho)=\mathrm{tr}[m_{\hat f}(L_{\rho}, R_{\rho})]-1.
\end{eqnarray}
The entropy $S$ quantifies the mixing of a $n$-dimensional quantum state $\rho$, with $S=0$ for pure states and $S=n-1$ for maximum mixedness. For a schematic distribution of predictability $P$, visibility $V$ and quantum $f$ entropy in quantum states $|i\rangle\langle i|$, Bell states, maximally coherent state and maximally mixed state, see FIG.\ref{fig0} and Appendix~\textbf{C}).

Combining Eqs.(\ref{equ}) and (\ref{f entropy}), we have the following trade-off relations.

\begin{theorem}\label{prop4}
For any $n$-dimensional state $\rho$ and $f\in \mathcal{F}_{op}$, we have the following complementary relation between the quantum $f$ entropy and the wave-particle duality.
\begin{eqnarray}\label{tradeoff}
P(\rho)+V(\rho)+\frac{1}{n-1}S_{f}(\rho)=1.
\end{eqnarray}
\end{theorem}

From the perspective of metric-adjusted skew information, Eq.(\ref{tradeoff}) can be further represented as
\begin{eqnarray}\label{pvif}
P(\rho)+V(\rho)=\frac{1}{n-1}\sum_{\alpha=1}^{n^2}I_{f}(\rho,X_{\alpha}),
\end{eqnarray}
where $\{X_{\alpha}:\alpha=1,2,\cdots,n^2\}$ constitutes an operator orthonormal basis. (\ref{pvif}) can be viewed as a complementary relation between the metric-adjusted skew information and the wave-particle duality.

Let $M_n(\mathbb{C})$ denote the set of all $n \times n$ complex matrices. For any $A \in M_n(\mathbb{C})$, density matrices $\rho_1, \rho_2$, and a function $f: [0, \infty) \rightarrow \mathbb{R}$, the quasientropy is defined by \cite{Hiai},
\begin{eqnarray}
S_f^A\left(\rho_1 \mid \rho_2\right)=\left\langle A \rho_2^{1 / 2}, f\left[\Delta\left(\rho_1 / \rho_2\right)\right]\left(\rho_2^{1 / 2}\right)\right\rangle,
\end{eqnarray}
where $\langle X, Y\rangle=\operatorname{tr}\left(X^{\dagger} Y\right)$ is the Hilbert-Schmidt inner product and $\Delta\left(\rho_1 / \rho_2\right): M_n(\mathbb{C}) \rightarrow M_n(\mathbb{C})$ refers to the linear mapping defined by $\Delta\left(\rho_1 / \rho_2\right)(X)=\rho_1 X \rho_2^{-1}$.
Since the quantum $f$ entropy and the quasientropy have the following relation \cite{Fan},
\begin{eqnarray}
S_f(\rho)=\sum_{\alpha=1}^{d^2} S_{\tilde{f}}^{X_\alpha}(\rho \mid \rho)-1,
\end{eqnarray}
a series of complementary relations can be established among the predictability, visibility and quasientropy,
\begin{eqnarray}\label{quasi}
P(\rho)+V(\rho)+\frac{1}{n-1}\sum_{\alpha=1}^{d^2} S_{\tilde{f}}^{X_\alpha}(\rho \mid \rho)=\frac{n}{n-1}.
\end{eqnarray}
In particular, for $n=2$ we have $P(\rho)+V(\rho)+\sum_{\alpha=1}^{d^2} S_{\tilde{f}}^{X_\alpha}(\rho \mid \rho)=2$.
As a direct consequence of the Theorem \ref{prop4} and the properties of quantum $f$ entropy, we have

\noindent{\bf Corollary.}
For any $n$-dimensional state $\rho$ and $f\in \mathcal{F}_{op}$, the corresponding predictability and visibility satisfy
\begin{eqnarray}\label{ineq}
P_{f}(\rho)+V_{f}(\rho)\leq 1.
\end{eqnarray}
\begin{table}
\renewcommand{\arraystretch}{2}
\centering
\caption{The expressions of function $f$, $m_{\hat{f}}$, predictability, visibility and quantum $f$ entropy related to Example 1.}
\resizebox{8.7 cm}{!}{
\begin{tabular}{cccccc}
\hline \hline $QFI$ & $f$  & $m_{\hat{f}}$ & $P_f$ & $V_f$ & $S_f$ \\
\hline
$\operatorname{WY}$ & $((1+\sqrt{x}) / 2)^2$ & $\sqrt{xy}$ & $1-\sqrt{1-r_3^2}$ & $\sqrt{1-r_3^2}-\sqrt{1-r^2}$  & $\sqrt{1-r^2}$\\
$\operatorname{SLD}$ & $(1+x) / 2$ & $2/(\frac{1}{x}+\frac{1}{y})$ & $r_3^2$ & $r^2-r_3^2$ & $1-r^2$ \\
& & &convex &convex  &concave\\
\hline \hline
\end{tabular}}
\end{table}

{\sf Remark} The quantum $(h,\phi)$ entropy $\mathbf{H}_{(h, \phi)}(\rho)$ of a state $\rho$ is defined by $\mathbf{H}_{(h, \phi)}(\rho)\equiv h[\operatorname{Tr} \phi(\rho)]$, where $h:\mathbb{R} \mapsto \mathbb{R}$ is a monotonic functional and $\phi:[0,1] \mapsto \mathbb{R}$ is continuous satisfying either (i) $\phi$ is strictly concave and $h$ is strictly increasing or (ii) $\phi$ is strictly convex and $h$ is strictly decreasing \cite{Bosyk}.
Denote $\phi(\rho)=m_{\hat{f}}\left(L_\rho, R_\rho\right)$. Although $P_{f}(\rho)$ in Eq.(\ref{predictability}) seems to be of the form of $h[\operatorname{Tr} \phi(\rho^d)]$ with $h(x)=\frac{n-x}{n-1}$, it fails to satisfy the conditions of a quantum $(h,\phi)$ entropy (it satisfies the convexity requirement on predictability). Nevertheless, when we consider the strictly increasing function $h(x)=\frac{x-1}{n-1}$, the quantum $(h, \phi)$ entropy $\mathbf{H}_{(h, \phi)}$ here gives rise to the following complementary relation with the predictability and visibility,
\begin{eqnarray}
P_{f}(\rho)+V_{f}(\rho)+\mathbf{H}_{(h, \phi)}(\rho)=1.
\end{eqnarray}
In conjunction with Theorem \ref{prop4}, the connection between quantum $f$ entropy and quantum $(h, \phi)$ entropy is also revealed in this way.

\begin{example}
In the computational basis $\{|0\rangle,|1\rangle\}$, we consider the qubit state in the Bloch representation:
$$
\rho=\frac{1}{2}\left(\textit{I}+\sum_{i=1}^3 r_i \sigma_i\right)=\frac{1}{2}\left(\begin{array}{cc}
1+r_3 & r_1-i r_2 \\
r_1+i r_2 & 1-r_3
\end{array}\right),
$$
where $r_i$ $(i=1,2,3) \in \mathbb{R}$, $\textit{I}$ is the identity operator, and $\sigma_1=\left(\begin{array}{ll}0 & 1 \\ 1 & 0\end{array}\right)$, $\sigma_2=\left(\begin{array}{cc}0 & -i \\ i & 0\end{array}\right)$ and $\sigma_3=\left(\begin{array}{cc}1 & 0 \\ 0 & -1\end{array}\right)$ are the Pauli matrices.
Let $r=\sqrt{r_1^2+r_2^2+r_3^2} \leqslant 1$ be the module of the Bloch vector of $\rho$. The eigenvalues of $\rho$ are $\lambda_1=\frac{1}{2}(1+r)$ and $\lambda_2=\frac{1}{2}(1-r)$. Meanwhile, the two eigenvalues of $\rho^d$ are given by
$\mu_1=\frac{1}{2}(1+r_3)$ and $\mu_2=\frac{1}{2}(1-r_3)$. Therefore, by direct calculation we have
\begin{equation}\label{exa1}
\begin{aligned}
& C_f(\rho)=\frac{2 r^2 f(0)}{3(1+r) f\left(\frac{1-r}{1+r}\right)}, \\
& V_f(\rho)=\frac{2 r^2 f(0)}{(1+r) f\left(\frac{1-r}{1+r}\right)}-\frac{2 r_3^2 f(0)}{\left(1+r_3\right) f\left(\frac{1-r_3}{1+r_3}\right)}, \\
& P_f(\rho)=\frac{2 r_3^2 f(0)}{\left(1+r_3\right) f\left(\frac{1-r_3}{1+r_3}\right)}, \\
& S_f(\rho)=1-\frac{2 r^2 f(0)}{(1+r) f\left(\frac{1-r}{1+r}\right)}.
\end{aligned}
\end{equation}
\end{example}

The pure states with $r_3=\pm1$ achieve the maximal visibility. In Table I we list the
formulas if the function $f$, the mean $m_{\hat{f}}$, the predictability, visibility and quantum $f$ entropy for qubit state $\rho$ in Example 1, where QFI, WY and SLD stand for quantum Fisher information, Wigner-Yanase skew information and symmetric logarithmic derivative \cite{Sun}, respectively.

\begin{figure}
\begin{center}
\includegraphics[scale=0.8]{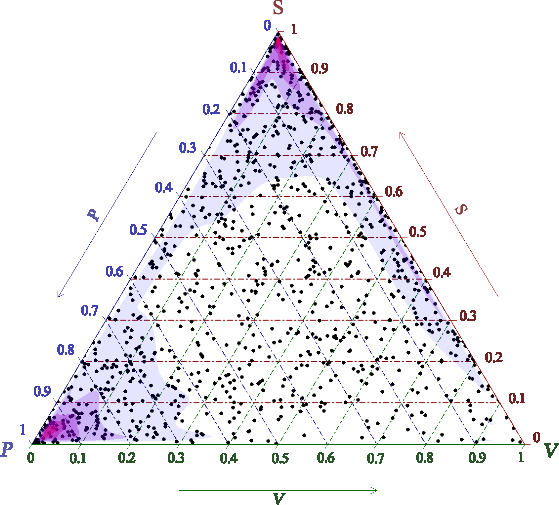}
\caption{Ternary phase diagram: predictability-visibility-entropy complementary for any qubit state with $f=f_{WY}$, where each data point on the graph represents the ternary trade-off relation of a distinct qubit state, with the 1000 qubit states being generated randomly.
}\label{fig1}
\end{center}
\end{figure}

Concerning the complementarity described by Eq.(\ref{exa1}) for any qubit state, FIG.\ref{fig1} shows the numerical results on the proportion of $P$, $S$ and $V$ distributions, which are randomly generated by the \emph{Mathematica} software.

\begin{figure*}
\begin{center}
\includegraphics[scale=0.7]{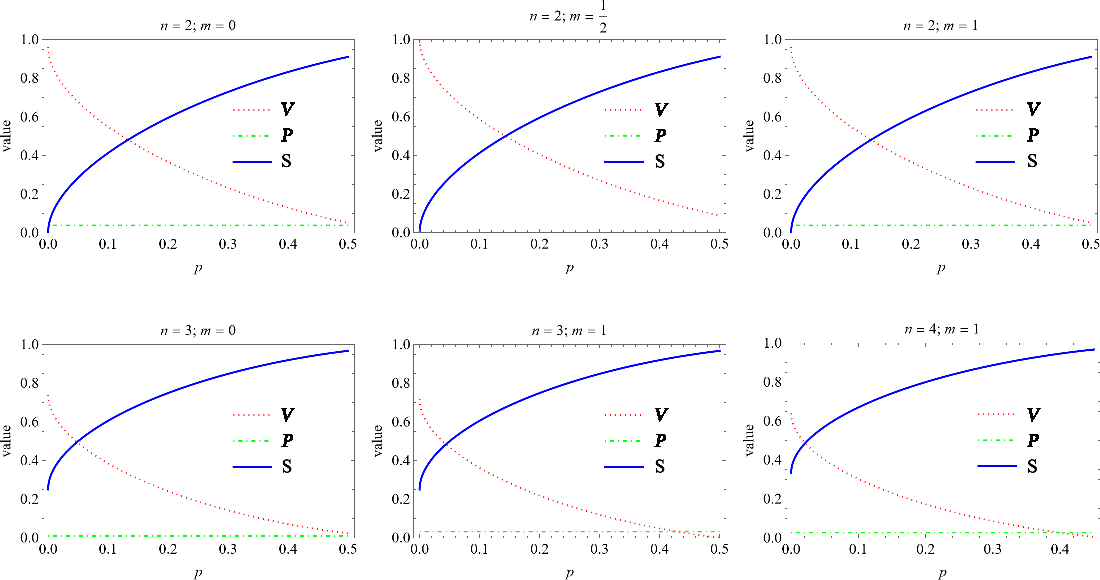}
\caption{The high-dimensional predictability-visibility-entropy complementary relationship of the Werner states as illustrated by Example 2. The behavior of $P$, $V$ and $\textbf{S}=\frac{1}{n^2-1}S_f$ varies with $p$ for 4-path, 9-path and 16-path for fixed $m$, respectively.
}\label{fig2}
\end{center}
\end{figure*}

\begin{example}\label{Exmp2}
We consider the Werner states
$$
W=\frac{n-m}{n^3-n} \textit{I}_n \otimes \textit{I}_n+\frac{nm-1}{n^3-n}\textit{F}, \quad m \in[0,1]
$$
on the $n^2$-dimensional Hilbert space $\mathbb{C}^n \otimes \mathbb{C}^n$, where $\textit{F}=$ $\sum_{\mu, \nu=1}^n|\mu\rangle\langle\nu|\otimes| \nu\rangle\langle\mu|$ is the flip operator with $\{|\mu\rangle: \mu=1,2, \ldots, n\}$ an orthonormal basis of $\mathbb{C}^n$. $\textit{W}$ has the following spectral decomposition \cite{Li},
$$
\textit{W}=\frac{2 p}{n^2+n} \Pi_s+\frac{2(1-p)}{n^2-n} \Pi_a, \quad p \in[0,\frac{1}{2}]
$$
with eigenvalues $\lambda_1=\frac{2 p}{n^2+n}, \lambda_2=\frac{2(1-p)}{n^2-n}$
of multiplicities $\left(n^2+n\right) / 2$ and $\left(n^2-n\right) / 2$, respectively. The eigenvalues of $\textit{W}^d$ are $\mu_1=\frac{m+1}{n^2+n}$ and $\mu_2=\frac{n-m}{n^3-n}$ of multiplicities $n$ and $n^2-n$, respectively. Here $\Pi_s (\Pi_a)$ is projection onto the symmetric (antisymmetric) subspace of $\mathbb{C}^{n^{2}}$, and $p=\operatorname{tr}\left(\textit{W} \Pi_s\right)$. We have
\begin{equation}\label{exa2}
\begin{aligned}
& C_{f_{\mathrm{WY}}}(\textit{W})=\frac{2n^2-\left(\sqrt{p\left(n^2+n\right)}+\sqrt{(1-p)\left(n^2-n\right)}\right)^2}{2(n^2+1)}, \\
& V_{f_{\mathrm{WY}}}(\textit{W})=\frac{n\left(\sqrt{(n-1)(n-m)}+\sqrt{m+1}\right)^2}{(n+1)(n^2-1)}\\
&\quad \quad \quad \quad
-\frac{\left(\sqrt{p\left(n^2+n\right)}+\sqrt{(1-p)\left(n^2-n\right)}\right)^2}{2(n^2-1)}, \\
& P_{f_{\mathrm{WY}}}(\textit{W})=\frac{n}{n^2-1}\left[n-\frac{\left(\sqrt{(n-1)(n-m)}+\sqrt{m+1}\right)^2}{n+1}\right], \\
& S_{f_{\mathrm{WY}}}(\textit{W})=\frac{1}{2}\left(\sqrt{p\left(n^2+n\right)}+\sqrt{(1-p)\left(n^2-n\right)}\right)^2-1.
\end{aligned}
\end{equation}

\end{example}

\begin{figure}
\begin{center}
\includegraphics[scale=0.9]{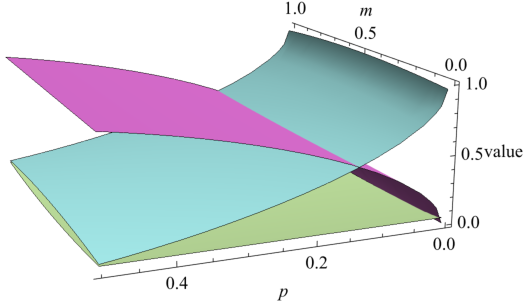}
\caption{Demonstration of the predictability-visibility-entropy complementarity of Werner states for $n=2$ with respect to $m$ and $p$. The concave surface (purple) above represents the normalized quantum $f$ entropy $\textbf{S}$ with respect to $p$. The upper convex surface (cyan) represents the visibility $V_{f}$ with respect to $p$ and $m$. The lower convex surface (grass green) represents the predictive $P_{f}$ with respect to $m$. Here $p \in[0,\frac{1}{2}]$ and $m \in[0,1]$.
}\label{fig3}
\end{center}
\end{figure}

Considering the complementarity of Werner states given by Eq.(\ref{exa2}), FIG.\ref{fig2} shows that the behavior of $P$, $V$ and $\textbf{S}$ varies with $p$ for high-dimensional states with fixed $m$. FIG.\ref{fig3} shows that the behavior of $P$, $V$ and $\textbf{S}$ varies with $p$ and $m$ for 4-path case ($n=2$). \\

\section{Conclusions}
Based on an arbitrary symmetric normalized regular operator monotone function $f\in \mathcal{F}_{op}$, we have constructed quantitative measures of particle and wave aspects respectively. Inherently, we have investigated the relationship between the wave-particle duality and the quantum coherence associated with the metric-adjusted skew information. We have also established the trade-off relations between the wave-particle duality and the quantum $f$ entropy. Our results reveal the profound relations among the predictability, visibility and quantum coherence, and may highlight further investigations on
relations between WPD and other quantum quantities such as entanglement and non-localities.

\section*{ACKNOWLEDGMENTS}
This work is supported by the National Natural Science Foundation of China (NSFC) under Grant No. 12075159 and No. 12171044, and the specific research fund of the Innovation Platform for Academicians of Hainan Province.

\appendix
\renewcommand{\appendixname}{Appendix~}
\renewcommand{\thesection}{\Alph{section}}
\renewcommand{\thesubsection}{\arabic{subsection}}
\renewcommand{\theequation}{S\arabic{equation}}
\renewcommand{\thefigure}{S\arabic{figure}}

\section{Proof of the Proposition~1}
\setcounter{equation}{0}
\renewcommand\theequation{A\arabic{equation}}
\label{sec:props1}
To prove proposition~1, consider the spectral decomposition of $\rho^d=\sum_{i}\lambda_{i}|i\rangle \langle i|$, $\sum_{i}\lambda_{i}=1$. In Ref.\cite{Hansen} it has been proven that the trace of $m_{\hat f}(L_{\rho^{d}}, R_{\rho^{d}})$ is given by
\begin{eqnarray}
\mathrm{tr}[m_{\hat f}(L_{\rho^{d}}, R_{\rho^{d}})]=\sum_{i,j}m_{\hat f}(\lambda_{i},\lambda_{j})\geq 1,
\end{eqnarray}
and the equality holds if and only if $\rho^d$ is a pure state.
Firstly, $P(\rho)=1$ is equivalent to $\mathrm{tr}[m_{\hat f}(L_{\rho^{d}}, R_{\rho^{d}})]=1$.
If there is one $i$ such that $\rho_{ii}=1$, which implies that $\rho^{d}$ has only one eigenvalue of 1 and the rest is 0. Thus $\rho^{d}$ is pure. The converse is obvious. Therefore $P(\rho)=1$ if and only if $\rho_{ii}=1$ for one $i$.

Secondly, $P(\rho)=0$ is equivalent to $\mathrm{tr}[m_{\hat f}(L_{\rho^{d}}, R_{\rho^{d}})]=n$. Let $\lambda_{i} (i=1,2,\cdots, k)$ be the nonzero eigenvalues of $\rho^{d}$ satisfying $\sum_{i}\lambda_i=1$. For any $f \in \mathcal{F}_{op}$, $
m_{\hat f}(\lambda_{i},\lambda_{j})
=\frac{1}{2}(\lambda_{i}+\lambda_{j}-(\lambda_{i}-\lambda_{j})^2 \frac{f(0)}{\lambda_{j}f(\frac{\lambda_{i}}{\lambda_{j}})})
\leq\frac{\lambda_{i}+\lambda_{j}}{2},
$
where the equality holds if and only if $\lambda_{i}=\lambda_{j}$. Hence,  $\sum_{i,j}m_{\hat f}(\lambda_{i},\lambda_{j})\leq \sum_{i,j}\frac{\lambda_{i}+\lambda_{j}}{2}\leq n$, and the last equality is saturated if and only if $\lambda_{i}=\lambda_{j}=\frac{1}{n}$. Thus $\mathrm{tr}[m_{\hat f}(L_{\rho^{d}}, R_{\rho^{d}})]=n$ if and only if $\lambda_{i}=\lambda_{j}=\frac{1}{n}$.

For item (3a), the permutation of the diagonal entries of density matrix $\rho$ (i.e., the eigenvalues of $\rho^{d}$) does not alter the value of $\sum_{i,j}m_{\hat f}(\lambda_{i},\lambda_{j})$.

Finally, it is known that if $f: \mathbb{R} \to \mathbb{R}$  is a continuous convex function, then the trace function $A\mapsto \mathrm{tr}\left(f\left(A\right)\right)$ is a convex function, see \cite{Klien}, and the combination of $m_{\hat f}(x,y)$ is an operator concave function proves that $-\mathrm{tr}[m_{\hat f}(x, y)]$ is convex. This concludes the proof.

\section{Proof of the Proposition~2}
\setcounter{equation}{0}
\renewcommand\theequation{B\arabic{equation}}
\label{sec:props2}
Here we provide the proof of Proposition~2. Firstly, we just need to prove that $\rho=\rho^{d}$ when $V(\rho)=0$.
We invoke generalized Klien's inequality \cite{Klien} on differentiable convex functions. For all Hermitian $n\times n$ matrices $A$, $B$ and all differentiable convex functions $f: \mathbb{R} \to \mathbb{R}$:
\begin{eqnarray}
\mathrm{tr}[f(A) - f(B)] \geq \mathrm{tr}[(A - B)f'(B)] .
\end{eqnarray}
If $f$ is strictly convex, the equality holds if and only if $A=B$. Combined with $A=\rho$ and $B=\rho^{d}$, this implies that
\begin{eqnarray}
(n-1)V(\rho)
&=&\mathrm{tr}[m_{\hat f}(L_{\rho^{d}},R_{\rho^{d}})-m_{\hat f}(L_\rho, R_\rho)]\nonumber\\
&\geq& \mathrm{tr}[(\rho^{d}-\rho)m_{\hat f}'(L_{\rho^{d}},R_{\rho^{d}})].
\end{eqnarray}
Therefore $V(\rho)=0$ if and only if $\rho=\rho^{d}$.

For item (2b), $V(\rho)=1$, which is equivalent to $\mathrm{tr}[m_{\hat f}(L_{\rho^{d}}, R_{\rho^{d}})-m_{\hat f}(L_\rho, R_\rho)]=n-1$, implies that $\rho$ is a pure state. This is because if $\rho$ is not a pure state, then $\rho$ has at least two different nonzero eigenvalues $\mu_1,\mu_2$. We have
$$
\mathrm{tr}[m_{\hat f}(L_{\rho^{d}},R_{\rho^{d}})]=n+\mu_{2}\hat f(\frac{\mu_1}{\mu_2})+\mu_{1}\hat f(\frac{\mu_2}{\mu_1})>n.
$$
 This is impossible. So $\mathrm{tr}[m_{\hat f}(L_{\rho}, R_{\rho})]=1$ and $\mathrm{tr}[m_{\hat f}(L_{\rho^{d}}, R_{\rho^{d}})]=n$. Therefore, $\lambda_{i}=\frac{1}{n}$ for all $i$. On the other hand, if $\rho$ is a pure state with equal diagonal elements, it is evident from the proof of Proposition~1 that $\mathrm{tr}[m_{\hat f}(L_{\rho^{d}}, R_{\rho^{d}})]=n$ and $\mathrm{tr}[m_{\hat f}(L_{\rho}, R_{\rho})]=1$. Thus $V(\rho)=1$.

 For item (3b), it is evident that $V(\rho)$ is invariant under permutations of the $n$ path labels.

 Finally, we need to show the convexity of $V(\rho)$. Considering the $2n$-dimensional quantum state $\rho=p_1 \rho_1 \oplus p_2 \rho_2$, $0\leq p_1,p_2\leq1$ and $p_1+p_2 =1$. We suppose $\mu_i$ and $\nu_j$ ($i,j=1,2,\cdots,n$) are eigenvalues of $\rho_1$ and $\rho_2$, then the eigenvalues of $\rho$ are $\{p_1 \mu_i,p_2 \nu_j\}_{ij}$. Thus,
\begin{eqnarray}
&\mathrm{tr}&[m_{\hat f}(L_{\rho}, R_{\rho})]\nonumber\\
&=&\sum_{i,j}m_{\hat f}(p_1\mu_{i},p_1\mu_{j})+\sum_{k,l}m_{\hat f}(p_2\nu_{k},p_2\nu_{l})\nonumber\\
&=&p_1\sum_{i,j}m_{\hat f}(\mu_{i},\mu_{j})+p_2\sum_{k,l}m_{\hat f}(\nu_{k},\nu_{l})\\
&=&p_1\mathrm{tr}[m_{\hat f}(L_{\rho_1}, R_{\rho_1})]+p_2\mathrm{tr}[m_{\hat f}(L_{\rho_2}, R_{\rho_2})],\nonumber
\end{eqnarray}
where the first and third equalities come from the definition of $\mathrm{tr}[m_{\hat f}(L_{\rho}, R_{\rho})]$. The second equality holds since $m_{\hat f}(tx,ty)=t m_{\hat f}(x,y)$ for any $t>0$. Similarly,
\begin{small}
$$\mathrm{tr}[m_{\hat f}(L_{\rho^d}, R_{\rho^d})]=p_1\mathrm{tr}[m_{\hat f}(L_{\rho_{1}^{d}}, R_{\rho_{1}^{d}})]+p_2\mathrm{tr}[m_{\hat f}(L_{\rho_{2}^{d}}, R_{\rho_{2}^{d}})].$$
\end{small}
Thus we have
\begin{eqnarray}\label{conv}
V(\rho)&=&V(p_1 \rho_1 \oplus p_2 \rho_2)\nonumber\\
&=&\frac{\mathrm{tr}[m_{\hat f}(L_{\rho^d}, R_{\rho^d})-m_{\hat f}(L_{\rho}, R_{\rho})]}{2n-1}\nonumber\\
&=&\frac{p_1\mathrm{tr}[m_{\hat f}(L_{\rho_{1}^d}, R_{\rho_{1}^d})-m_{\hat f}(L_{\rho_{1}}, R_{\rho_{1}})]}{2n-1}\nonumber\\
&+&\frac{p_2\mathrm{tr}[m_{\hat f}(L_{\rho_{2}^d}, R_{\rho_{2}^d})-m_{\hat f}(L_{\rho_{2}}, R_{\rho_{2}})]}{2n-1}\nonumber\\
&=&\frac{n-1}{2n-1}[p_1V(\rho_{1})+p_2V(\rho_{2})].
\end{eqnarray}
Suppose the $\rho_1,\rho_2$ are states of a system $A$. We introduce an auxiliary system $B$ whose state space has an orthonormal basis $|0\rangle$ and $|1\rangle$. Assume that the initial joint state of $AB$ is:
$$
\rho^{AB}=p_1\rho_1\otimes|0\rangle\langle0|+p_2\rho_2\otimes|1\rangle\langle1|,
$$
where $0\leq p_1,p_2\leq1$ and $p_1+p_2 =1$. We consider an incoherent operation $\Phi^{AB}$ on the whole system $AB$ with Kraus operators $\left\{K_i\otimes|0\rangle\langle j|\right\}, i=1,2,\cdots,n, j=0,1$. Then $\Phi^{AB}(\rho^{AB})=\sum_{ij}K_{i}\otimes|0\rangle\langle j|(\rho^{AB})(K_i\otimes|0\rangle\langle j|)^\dag$. We have \cite{Nielsen2000}
\begin{eqnarray}
\Phi^{AB}(\rho^{AB})=(p_1 \rho_1 + p_2 \rho_2)\otimes|0\rangle\langle 0|.
\end{eqnarray}
By definition we can easy to know that
\begin{eqnarray}
V(\Phi^{AB}(\rho^{AB}))&=&V((p_1 \rho_1 + p_2 \rho_2)\otimes|0\rangle\langle 0|)\nonumber\\
&=&\frac{n-1}{2n-1}V(p_1 \rho_1 + p_2 \rho_2).
\end{eqnarray}
Due to Eq.(\ref{conv}) and the monotonicity of $V(\rho)$ under incoherent operation, we obtain
\begin{eqnarray}
V(p_1 \rho_1 + p_2 \rho_2)&=&\frac{2n-1}{n-1}V(\Phi^{AB}(\rho^{AB}))\nonumber\\
&\leq& \frac{2n-1}{n-1}V(\rho^{AB})\nonumber\\
&=&\frac{2n-1}{n-1}V(p_1 \rho_1\oplus p_2 \rho_2)\nonumber\\
&=&p_1V(\rho_{1})+p_2V(\rho_{2}),
\end{eqnarray}
from which item~(4b) follows.

\section{Basic properties of quantum $f$ entropy}
Quantum $f$ entropy $S_f(\rho)$ defined by Eq.(\ref{f entropy}) has the following properties, which can be directly verified \cite{Fan}.
\begin{enumerate}
\item[(i)] $S_f(\rho)$ is non-negative. The quantum $f$ entropy is zero if and only if the state $\rho$ is pure.
\item[(ii)] In a $n$-dimensional Hilbert space, the quantum $f$ entropy is at most $n-1$. The $f$ entropy is equal to $n-1$ if and only if the system is in the completely mixed state $\textit{I}/n$.

\item[(iii)] Suppose a composite system $AB$ is in a pure state $|\psi^{AB}\rangle$, then $S_f\left(\rho^A\right)=S_f\left(\rho^B\right)$, where $\rho^A=$ $\operatorname{tr}_B|\psi\rangle\langle\psi|$ and $\rho^B=\operatorname{tr}_A|\psi\rangle\langle\psi|$.
\item[(iv)] $S_f(\rho)$ is concave, i.e.,
$S_f(\sum_j p_j \rho_j) \geq \sum_j p_j S_f\left(\rho_j\right)$,
where $p_j$ are probabilities.
\item[(v)] $S_f(\rho)$ is unitary invariant, i.e., $S_f(U \rho U^{\dagger})=S_f(\rho)$.
\item[(vi)] $S_f(\rho \otimes \textit{I} / n)=S_f(\textit{I} / n \otimes \rho)=n S_f(\rho)+S_f(\textit{I} / n)$.
\item[(vii)] $S_f\left(\sum_j p_j|j\rangle\langle j| \otimes \rho_j\right) \geq \sum_j p_j S_f\left(\rho_j\right)$, where $p_j$ are probabilities, $\{|j\rangle\}$ are orthogonal states for a system $A$, and $\rho_j$ are quantum states on another system $B$.
\end{enumerate}


\end{document}